# Atomistic Simulations of Nanotube Fracture

T. Belytschko[1], S.P. Xiao[2], G.C. Schatz[3] and R. Ruoff[4]

Department of Mechanical Engineering
2145 N Sheridan Rd
Northwestern University
Evanston, IL 60208

ABSTRACT

The fracture of carbon nanotubes is studied by atomistic simulations. The fracture behavior is found to be almost independent of the separation energy and to depend primarily on the inflection point in the interatomic potential. The range of fracture strains compares well with experimental results, but the predicted range of fracture stresses is markedly higher than observed. Various plausible small-scale defects do not suffice to bring the failure stresses into agreement with available experimental results. As in the experiments, the fracture of carbon nanotubes is predicted to be brittle. The results show moderate dependence of fracture strength on chirality.

1. INTRODUCTION

We report here molecular mechanics and molecular dynamics studies of the failure of nanotubes. Nanotubes, since their discovery in 1991[1], have attracted much interest because of their ability to sustain large deformations and their great stiffness and possible high strength. The ability of carbon nanotubes to sustain large bending deformations has been observed experimentally and verified by molecular dynamics studies[2][3][4]. Molecular dynamics studies have also replicated the ability of nanotubes to sustain very distorted configurations with only elastic deformations and no creation of atomic defects[2][3][5][6].

Previous molecular mechanics simulations predict that ideal single-walled carbon nanotubes should possess extremely high tensile strengths. The strain at tensile failure has been predicted to be as high as ~30% for SWCNTs[7] by molecular mechanics. However, in the only experiments[8] reported so far, the tensile failure strain is usually between ~10% to ~13% and as low as 2%. A recent study[9] of SWCNTs with the tight binding approximation describes a mechanism of defect nucleation in which two hexagon pairs are converted, through a Stone-Wales bond rotation, into 5/7/7/5 (pairs of pentagon-heptagon) defects. It was hypothesized that at high temperatures, plastic response may occur due to separation and glide of these 5/7/7/5 defects whereas at lower temperature this may result in fracture[9][10][11].

Molecular simulations of fracture are now quite commonplace. The earliest atomistic simulations of fracture were reported in the early 1970's[12][13][14] and many materials have been investigated. Works closely related to this are Shenderova et al[15] who examined the fracture of polycrystalline diamond, and Omeltchenko et al[16] who studied notched graphene.

---

1. Walter P Murphy Professor and Chair of Mechanical Engineering, tedbelytshko@northwestern.edu
2. Research Assistant
3. Professor of Chemistry
4. Professor of Mechanical Engineering

In this paper, the fracture of carbon nanotubes is studied by molecular mechanics and dynamics. We show that the experimental results of Yu et al[8] for the fracture strain and stress of carbon nanotubes can be reproduced reasonably well by molecular mechanics. Furthermore, we show that some of the variability of failure strains reported by Yu et al[8] can be partially explained by the presence of defects and variations due to chirality. The simulations always exhibit brittle fracture, which also agrees with these experiments.

Nanotubes are particularly attractive for studying fracture by molecular mechanics because they are single molecules with all atoms joined by identical bonds. Thus the degree of heterogeneity and variety of length scales found in fracture of most materials is absent in nanotubes. The major defects which have been postulated are the previously mentioned Stone-Wales defects that are of the order of several bond lengths. In experiments performed in the TEM environment, it is possible for single atoms to be ejected by impact, see Smith[17] and Barnhart[18], but there are also small-scale defects.

Of course, the interpretation of fracture simulations by atomistic models must be treated with care, since the reconfiguration of bonds is not treated. However, we show that for small initial defects, the fracture strength depends primarily on the inflection point of the interatomic energy and is almost independent of the dissociation energy. Since the inflection strain occurs substantially before the strain associated with of bond breaking, where the formation of other bonds is expected, these results may give the correct picture of fracture at moderate temperatures ($0 \sim 500\,K$).

2. COMPUTATIONAL METHODS

We used standard molecular mechanics and molecular dynamics methods. By molecular mechanics, we refer to methods where the equilibrium configuration of the model system is sought by minimizing the energy, which consists of the sum of the interatomic potentials minus any work by external forces. Such methods imply a temperature of $0\,K$ and cannot account for the effects of temperature. In molecular dynamics methods, the momentum equations are integrated in time for the system of atoms with interatomic forces given by the interatomic potential. Thus the effects of nonzero temperatures are included, although there are always limitations of such predictions due to the statistical nature of molecular motion for systems of bonded atoms.

The interatomic bonds in nanotubes are hybridized $sp^2$ bonds. In most of our calculations, the interatomic potential used is a modified Morse potential function in which a bond angle-bending is added

$$E = E_{stretch} + E_{angle} \qquad (1)$$

$$E_{stretch} = D_e \{[1 - e^{-\boldsymbol{b}(r-r_0)}]^2 - 1\} \qquad (2)$$

$$E_{angle} = \frac{1}{2} k_{\boldsymbol{q}} (\boldsymbol{q} - \boldsymbol{q}_0)^2 [1 + k_{sextic} (\boldsymbol{q} - \boldsymbol{q}_0)^4] \qquad (3)$$

where $E_{stretch}$ is the bond energy due to bond stretch, and $E_{angle}$ is the bond energy due to bond angle-bending, $r$ is the length of the bond, and $\boldsymbol{q}$ is the current angle of the adjacent bond, a standard deformation measure in molecular mechanics. The parameters are

$$r_0 = 1.39 \times 10^{-10}\,m,\ D_e = 6.03105 \times 10^{-19}\,N \cdot m,\ \boldsymbol{b} = 2.625 \times 10^{10}\,m^{-1}$$

$$\boldsymbol{q}_0 = 2.094\,rad,\ k_{\boldsymbol{q}} = 0.9 \times 10^{-18}\,N \cdot m/rad^2,\ k_{sextic} = 0.754\,rad^{-4}$$

This choice of parameters corresponds to a separation (dissociation) energy of 124Kcal/mole (5.62eV/atom). This is the usual Morse[19] potential except that the angle-bending energy has been added and the constants are slightly modified so that it corresponds with the Brenner potential for strains below 10%.

The bond angle-bending potential does not contribute to the stretching energy. It is added to stabilize the molecular structure in the tubular configuration. Without a bond angle-bending energy, a stable configuration cannot be found for the nanotube. Instead, it tends to collapse onto itself. So the bond angle potential plays an essential role in establishing an equilibrium configuration of the nanotube; however, it has little effect on fracture.

The Brenner function[19] is considered more accurate and is more versatile (it can handle bond hybridization and bonds with atoms other than carbon) than the Morse potential for carbon. It is given by:

$$E_{ij} = \frac{1}{2} \sum_{j(\neq i)} [V_R(r_{ij}) - \overline{B}_{ij} V_A(r_{ij})] \qquad (4)$$

where $\overline{B}_{ij}$ is a function of bond angle and

$$V_R(r_{ij}) = f(r_{ij}) \frac{D_e}{S-1} e^{-\sqrt{2S}\,b(r-r_e)} \qquad V_A(r_{ij}) = f(r_{ij}) \frac{D_e S}{S-1} e^{-\sqrt{\frac{2}{S}}\,b(r-r_e)}$$

The function $f(r_{ij})$, which restricts the pair potential to nearest neighbors, is given by

$$f(r) = \begin{cases} 1 & r < r_1 \\ \frac{1}{2}\{1 + \cos[\frac{p(r-r_1)}{r_2 - r_1}]\} & r_1 \leq r \leq r_2 \\ 0 & r > r_2 \end{cases} \qquad (5)$$

where

$$r_0 = 1.4507 \times 10^{-10} m,\ D_e = 9.648 \times 10^{-19} N \cdot m,\ b = 2.1 \times 10^{10} m^{-1}$$
$$S = 1.22,\ r_1 = 1.7 \times 10^{-10} m,\ r_2 = 2.0 \times 10^{-10} m$$

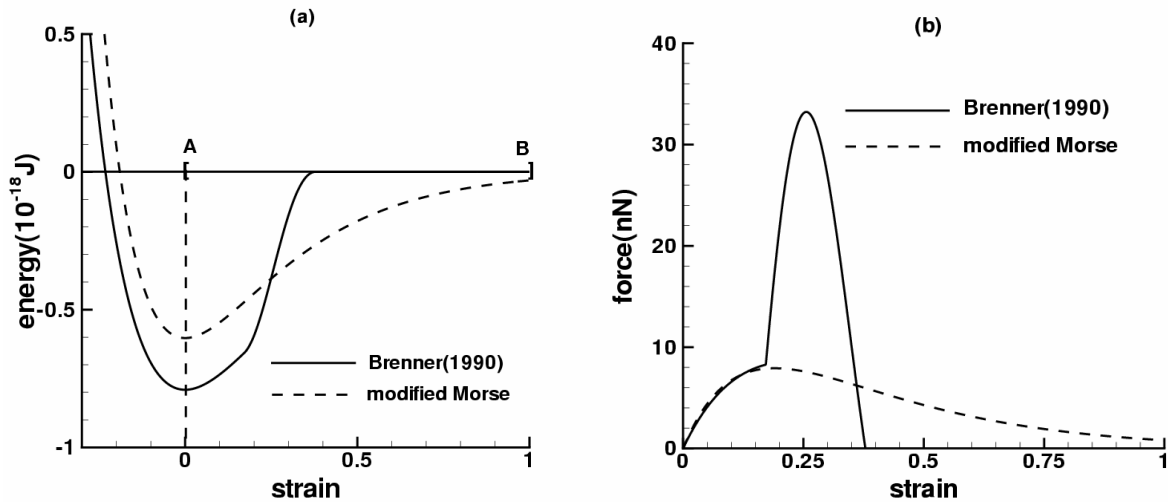

Fig. 1. The Brenner and modified Morse potentials and tensile force fields.
(a) potential field for Brenner and modified Morse potential (b) force fields in segment [AB] in part (a)

Fig. 1 compares the interatomic stretching energy and force for the Brenner and Morse potentials in the tensile regime. Because the stretching energy dominates the behavior in fracture, we show the relationships between force and bond strain with the bond angle kept constant. Strain is denoted by $e$ and is defined by $e = \frac{l - l_0}{l_0}$ where $l_0$ is the equilibrium (initial) length and $l$ the current length of the bond (i.e. the distance between nuclei). In the Brenner potential, the switch function $f(r_{ij})$ introduces a dramatic increase in the interatomic force at $r = r_1$ (the camel-back on the force curve), which rises sharply with a peak at around 30% strain. This strange feature in the force is a result of the cutoff in the switching function on the interatomic potential. Shenderova, et al[15] also noted this behavior and shifted the switching function to larger strains (specific values are not given) so that it occurs after the inflection point in the interatomic potential. We have found that the cutoff affects fracture behavior even when it is shifted to 100% strain.

The average Young's modulus and Poisson's ratio $n$ for the nanotube with the modified Morse potential function in the range of 0 to 10% strain are $0.94\,TPa$ and $0.29$. They are comparable to those of the Brenner potential function which are $1.07\,TPa$ and $0.19$[3]. The nanotube is quite nonlinear for strains on the order of 10%. Therefore, for preciseness we specify both the tangent modulus at 0% strain and the secant moduli. The tangent modulus is $1.16\,TPa$, the secant moduli at 5% and 10% strains are $0.89\,TPa$ and $0.77\,TPa$, respectively. Van Lier et al[20] obtained $E = 1.14\,TPa$, $v = 0.11$ with a Hartree-Fock 6-31G* calculation on a 150 atom [9,0] nanotube. Zhou et al[21] obtained $E = 0.764\,TPa$ and $v = 0.32$ with local density calculations. Salvetat et al[22] measured $E = 0.81 \pm 0.41\,TPa$. Thus our value falls in the range of these computed and experimented values. It is noteworthy that our calculations show large differences between the tangent modulus and the secant moduli, it is only possible to measure secant moduli with moderate precision, so some of the experimental scatter may be due to different strain levels. The separation (dissociation) energy of this Morse potential is 124kcal/mole, which corresponds to the benzene carbon-carbon bond.

Certain qualitative and quantitative features of the interatomic force curve play key roles in fracture behavior. The strain at which the tensile force achieves its peak will be called the inflection strain since it corresponds to an inflection point in the potential. The interatomic force at this point will be called the peak force. The strain at which the tensile force vanishes is called the bond-breaking strain. The separation or bond-breaking energy (also called the dissociation energy) is the area under the force-strain curve (multiplied by the initial length), and it also corresponds to the change in the potential.

3. RESULTS

In the experiments of Yu et al[8], arc-grown multi-walled nanotubes were used. The experimental setup is shown in Fig. 2. The multi-walled nanotubes are attached to opposing AFM cantilever tips by a solid carbonaceous material deposited onto the tips. In most cases, only the outer nanotube was attached to the cantilever. The multi-walled nanotubes failed in a sheath-like pattern with typically only the outer nanotube failing. Therefore, only the outer nanotube was modeled in these studies. Any interactions of the outer nanotube with the inner tubes were neglected.

The outer nanotube in the experiments of Yu et al[8] varied in length from $1.80\,nm$ to $10.99\,nm$ and their diameters varied from $13\,nm$ to $36\,nm$; typically the multi-walled nanotubes consisted of 2 to 19 nested nanotubes. The smallest of the outer nanotubes had about 180 atoms around the circumference, and 24,000 atoms along the length, which corresponds to a total of 4.3 million atoms. The largest outer nanotube consisted of approximately 53.8 million atoms. Treating so many atoms would entail very bng computations, so most of our studies were conducted on smaller models. In the initial simulations, a [20, 0] zig-zag tube (the tube is labeled by the usual pair of integers $[n_1, n_2]^{[23]}$), consisting of 840 atoms were considered. The dimensions were: radius: $0.76 nm$ and length: $4.24 nm$. We also studied larger nanotubes to show that these results are essentially independent of the size for the defect-free nanotube; these are described later.

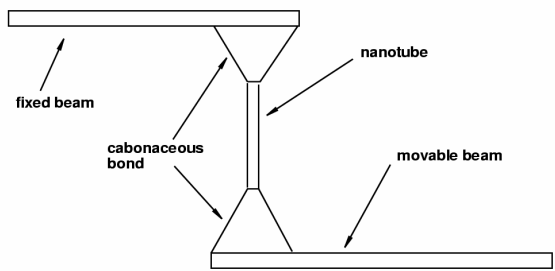

Fig. 2. An individual MWCNT mounted between two opposing AFM tips

We first considered molecular mechanics simulations. A single bond at the center of the tube was weakened by 10% in all molecular mechanics simulations except where noted. The motivation for weakening one bond is that the nanotube is a multi-atom molecule and few defects are known that could serve as a nucleation site for the crack (except for the Stone-Wales dislocation[4] which we consider later). Since the nanotube is a single molecule, there is no reason for weakening one bond if we were only interested in zero-temperature behavior. However, our objective is to ascertain room temperature behavior where the velocities of the atoms are random and nonzero. The kinetic energy of these atoms follows a Boltzmann distribution, and thus as the potential energy nears the inflection point due to the applied force, some bonds will be stretched beyond the point of maximum force. We mimic this by the 10% imperfection.

One end of the nanotube was axially displaced so that the nanotube was stretched. The applied force is computed by summing the interatomic forces for the atoms along the edge where the displacement is prescribed. Instead of reporting a deflection and force, we report strain and a nominal stress. The strain is given by $e = \dfrac{L - L_0}{L_o}$, where $L_0$ and $L$ are the initial and current length of the nanotube, respectively. The stress is calculated from the cross-sectional area $S$ ($S = \pi d h$, where $d$ is the diameter of the nanotube and $h$ is chosen as the interlayer separation of graphite, 0.34nm).

For the Brenner potential, the nanotube in the numerical simulation failed at an elongation of 28%[2] and the tensile strength was $110\ GPa$. This agrees with the results of Campbell et al[7] but does not agree with the Yu et al[8] experiments, where most of the samples failed between 10% and 13% strain (one is as low as 2%) and the tensile strengths range from $11\ GPa$ to $63\ GPa$. From our studies, we inferred that the high fracture strain can be attributed to the camel-back in the Brenner potential due to the cutoff in the switching function mentioned in the previous section.

With the modified Morse potential function, these numerical simulations give a failure strain of 15.7% and a failure stress of 93.5 $GPa$. The normalized stress-strain results along with the experimental results of Yu et al[8] are shown in Fig. 3. As can be seen, the failure strain is somewhat greater than the highest experimental value, while the failure stress is significantly higher. The computed stress-strain curve exhibits a sudden drop in stress at failure, so the fracture can be considered brittle. Yu's experiments also show brittle fracture, for the reported force continues to increase until the final measurement. Young's modulus for these experimental records is significantly below those previously reported (see beginning of this section). It is possible that some slippage occurred at the attachments for the high strain cases reported in Yu et al[8], resulting in a decrease in the measured values of Young's modulus.

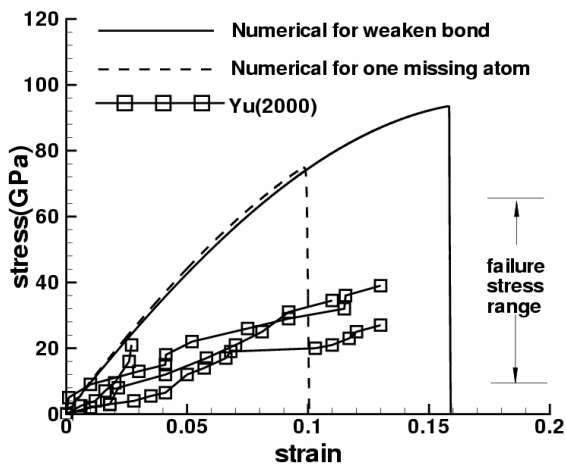

Fig. 3. Force-deflection curve for a model of zig-zag nanotube as compared to experimental results (normalized to stress vs. strain)

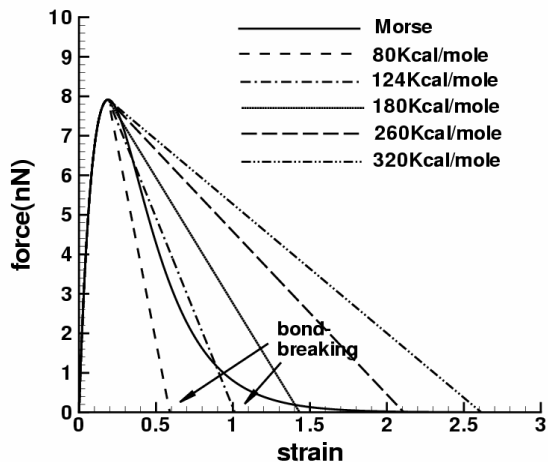

Fig. 4. Interatomic force-strain relationships for different separation energies for modified Morse potential

To get better insight into the relationship between fracture and separation energy, we varied the separation energy for the Morse function but kept the inflection point (i.e the maximum of the interatomic force) unchanged. For this purpose, we used linear approximations to the Morse force field after 19% strain as shown in Fig. 4. Often it has been noted that potential functions are quite well described using universal functions such as Rydberg[24][25] and Morse[26] functions, for which the location of the inflection point and the separation energy are strongly connected. This means that realistic potential functions are a good deal less flexible than we have assumed in Fig. 4. However, although there may be some relationship between separation energy and the inflection point, we doubt it is as strong as indicated in the preceding, for there would then be no need for first principle models.

Table 1 Failure strains of [20,0] nanotube for modified Morse potentials with various separation energies and interatomic force peaks at 19% and 13% strain

| Separation energy(Kcal/mole) | Failure strain of tube for Inflection at 19% strain | Failure strain of tube for Inflection at 13% strain |
| --- | --- | --- |
| 80 | 15.0% | 10.3% |
| 100 | 15.1% | 10.4% |
| 124 | 15.2% | 10.6% |
| 150 | 15.4% | 10.7% |
| 180 | 15.6% | 10.8% |
| 220 | 15.7% | 11.0% |
| 260 | 15.9% | 11.1% |
| 320 | 16.1% | 11.4% |

Table 1 gives the failure strains of the nanotube for different separation energies. As before, one end was displaced axially to stretch the nanotube. In addition to the Morse potential in which the inflection point (i.e. the peak force) occurs at 19% strain, we considered a potential with the inflection point at 13% strain. The temperature was considered to be $0K$. It can be seen from Table 1 that the failure strain depends very little on the separation energy but is strongly dependent on the location of the inflection point. The studies for 19% inflection point were repeated for a [60,0] nanotube (separation energy of 124 Kcal/mole) to check the effect of nanotube size. The failure strain and stress were almost identical. Fig. 5 shows the evolution of the crack in the tube. It can be seen that the bond-breaking spreads sideways after the initially weakened bond fails.

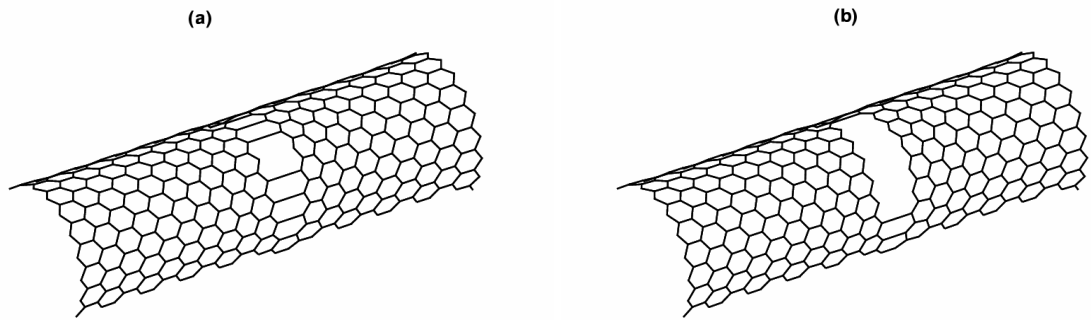

Fig. 5. Evolution of the crack in the nanotube

According Table 1, the failure strains are about 15.5% for a large range of separation energies with the inflection at 19% strain. This failure strain of the nanotube is smaller than the strain at which a single interatomic force achieves its peak, 19%. This can be explained by the hexagonal molecular structure of the nanotube. As a consequence of the structure of the bonds, the strains in the longitudinal interatomic bonds are larger than the average strain in the nanotube. For an inflection point of 13%, the failure strains are around 10.8%. The ratio of the failure strain to the inflection strain (10.6%/13%) is almost identical to the other case (15.2%/19%). The simulations show that the failure strain of the nanotube is essentially independent of the separation energy and depends primarily on the inflection point of the interatomic potential.

We have considered a large range of separation energies because the potential energy function is highly uncertain for strains beyond the inflection point, and we want to show that the results are not dependent on the separation energy. The simple interatomic potential used here is certainly not capable of describing the proper behavior of the system when bonds are broken, and indeed, breaking of the $sp^2$ bond in a carbon nanotube can result in a variety of phenomena that go beyond what our model can do. For example, new bonds should form between the dangling orbital in the carbons at the edge of the fracture region[27], probably with rehybridization, and structural transformations like the Stone-Wales transformation can take place in rings near the edge so as to seal the end of the fractured tube. Some of these phenomena will depend on the surroundings of the carbon nanotube, and temperature is also expected to play a role. Thus if fracture strength depended strongly on the separation energy, it would not be reasonable to model fracture with molecular mechanics. Fortunately, our results show a very weak dependence of fracture strain on the separation energy, so the shape of the potential surface after the inflection point cannot be important to fracture behavior.

As mentioned in the introduction, single atom may be ejected from the nanotube in the TEM environment, so we simulated this situation by removing a single atom and the three bonds, which include this atom. For the nanotube with one missing atom at the center, the failure strain is 10% while failure stress is $74\,GPa$. For the standard 124Kcal/mole separation energy, the normalized force-deflection curve for the zig-zag nanotube with one missing atom is shown in Fig. 3. The force-deflection curve is almost unchanged until fracture. However, fracture occurs at a much lower stress and strain. The fracture strain is now below the most commonly observed values, whereas the fracture stress is still greater than all observed values.

We next examine the effect of the chirality on the fracture strain. Yu et al[8] were not able to measure the chirality of the outermost nanotubes, so we cannot compare the effect to experiments. We ran three simulations in addition to the zig-zag nanotube:
1. [12,12] armchair nanotube
2. [16,8] chiral nanotube
3. [16,4] chiral nanotube

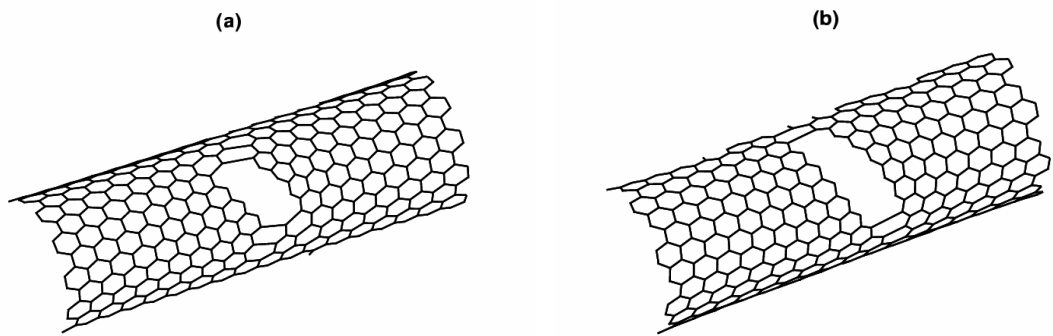

Fig. 6. Crack evolution in the (a) [12,12] armchair nanotube and (b) [16,8] chiral nanotube

The modified Morse potential with the separation energy of 124 Kcal/mole was used. Molecular mechanics simulations with the motion of one end displacement prescribed as before were made.

Fig. 6 shows the crack evolution for the nanotube, showing that the bonds nearest to parallel with the axis are the ones that break. The computed force-deflection curves are shown in Fig. 7. It can be seen that the calculations predict a significant dependence on the chirality of the nanotube on the strength. The computed failure strain of the [12,12] armchair nanotube is 18.7%, and the failure strength is $112\ GPa$. Thus the armchair nanotube is stronger than the zig-zag nanotube, and the strength of the [16,8] nanotube (17.1% failure strain and $106\ GPa$ failure stress) is between the two.

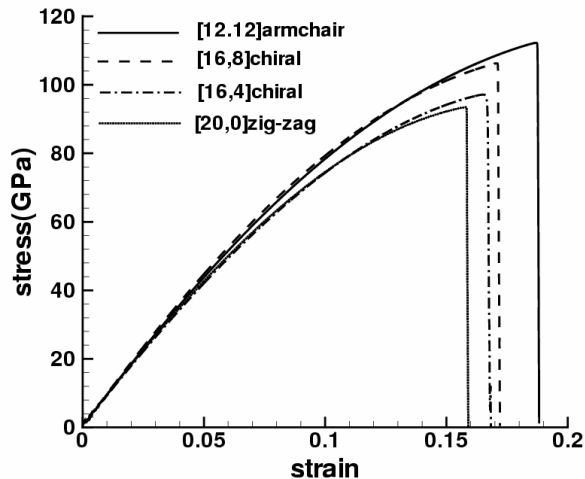

Fig. 7. Normalized force-deflection curves for nanotubes of different chiralities

In order to ascertain the appropriateness of weakening a bond in the molecular mechanics studies, we made several molecular dynamics simulations without any weakened bonds. For a temperature $T$, the thermal energy is given by:

$$E_{thermal} = C_p T \qquad (6)$$

where $C_p = 8.64\ J\ mol^{-1} K^{-1}$. A Boltzmann distribution is used for the velocities which are related to the thermal energy by

$$E_{thermal} = \frac{1}{2} \sum_{I=1}^{N} \sum_{i=1}^{3} m_I v_{iI}^2 \qquad (7)$$

where $N$ is the numbers of atoms, $m_I$ is the mass of atom $I$ and $v_{iI}$ are the components of the velocity of atom $I$.

The modified Morse potential with an inflection point at 19% strain and 124Kcal/mole separation energy was used to make a simulation at a temperature of $300\ K$. The Yu et al[8] experiments were completely static in a mechanical sense: the loads were applied over several minutes while the lowest period of the free-free nanotube (length = $4.24\ nm$) is estimated to be $0.21\ ps$. In molecular dynamics simulations of essentially static phenomena, a common hazard is that if the loading (or prescribed displacement) is applied too quickly, stress wave-like phenomena occur with considerable overshoot in interatomic forces. These can lead to spurious fracture.

Applying the prescribed displacement from zero to the failure strain in a molecular dynamics simulation without inducing spurious fracture is proved to be too expensive. Therefore we ran the simulation as follows:
1. a molecular mechanics simulation was used to bring the nanotube to 12% tensile strain.
2. the velocities due to temperature were applied to the atoms and the model was brought to steady-state by integrating the momentum equations for the atoms in time (steady-state is often called equilibrium).
3. the displacements at the end were prescribed to provide an extensional strain rate of 0.0072/ps.

In molecular dynamics simulations, the location of the crack was quite arbitrary and varied from simulation to simulation, but the scatter in the fracture strength was quite small. The nanotube failed at 16.1% ± 0.2% strain at a stress of 94 $GPa$. This compares quite well with the 15.8% strain and 93.5 $GPa$ stress obtained from the molecular mechanics simulation with a 10% weakened bond.

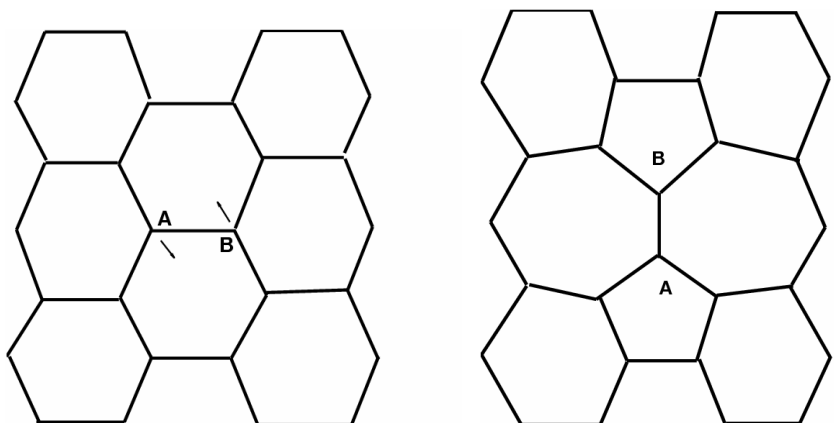

Fig. 8. Defect in (a) before rotation and (b) after rotaion

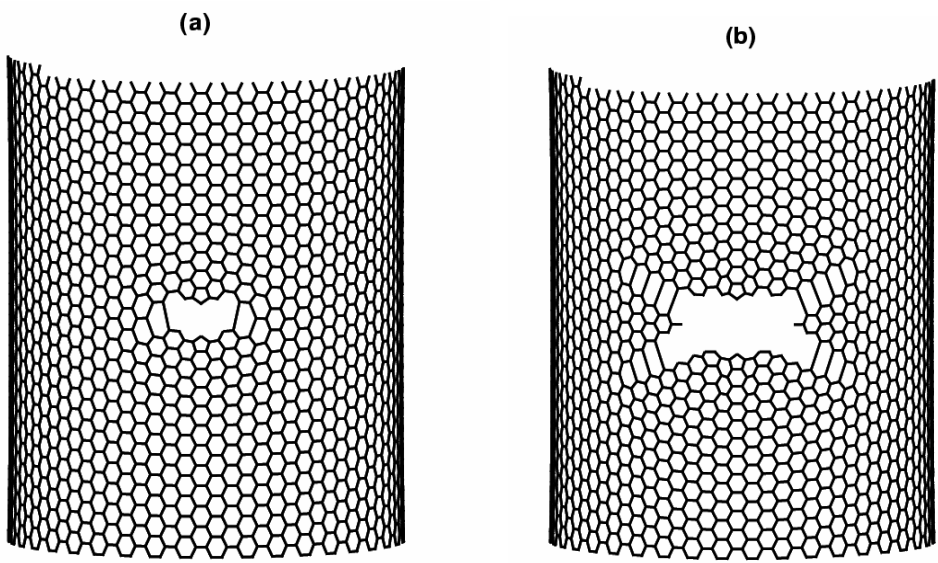

Fig. 9. Crack formation in the [40,40] armchair nanotube with 5/7/7/5 defect at (a) 12.7ps and (b) 12.8ps

One topological defect which has recently been identified theoretically is the 5/7/7/5 defect[10]. This defect corresponds to a 90 deg rotation of a C-C bond about its center as shown in Fig. 8, and is called a Stone-Wales transformation[4]. It results in two pentagons and two heptagons coupled in pairs. To study the effect of this defect on the behavior of the nanotube, we simulated a single defect at the center of the [12,12] armchair nanotube by molecular dynamics. The displacement was prescribed at the end of the nanotube and the 124 Kcal/mole modified Morse potential was used. The failure strain is 14.3% and the failure stress is 97.5 $GPa$. Compared to the armchair nanotube with a weakened bond shown in Fig. 7, the 5/7/7/5 dislocation gives a somewhat lower failure strain and lower tensile strength. The fracture is still brittle.

The study of the 5/7/7/5 defect was repeated with an [40,40] armchair nanotube. In this case the failure strain of the tube is 14.2% and the fracture is still brittle. So the size of the nanotube has little effect on the strength. Fig. 9 shows that for the 5/7/7/5 dislocation, the maximum shear strain occurs in the $\pm \frac{\boldsymbol{p}}{4}$ direction[10]. Thus the cracks grow in the direction of maximum shear.

## 4. CONCLUSIONS

It has been shown that some salient features of fracture behavior, i.e its brittle character and the magnitudes of fracture strains, of carbon nanotubes can be explained by atomistic simulations. In these simulations, the ranges of failure strains correspond roughly to observed values. However, the predicted fracture stresses are too large, even when defects are included. It was shown that the fracture strain of extremely homogeneous systems with covalent bonds, such as the carbon nanotube, appears to be driven primarily by the inflection point in the interatomic potential. The separation (dissociation) energy has little effect on the fracture strain.

Various nanotube chiralities were studied. Chirality appears to have a moderate effect on the strength of nanotubes, with the failure stress varying from 93.5 $GPa$ to 112 $GPa$ and the failure strain from 15.8% to 18.7%. In all cases the fracture was predicted to be brittle, which agrees with experiments.

Because of the homogeneity of multi-atom molecules such as the nanotube, defects for nucleating a fracture are not immediately apparent. To provide a nucleation site for a crack in molecular simulations, we reduced the strength of a single bond. Our motivation is that when the nanotube is near failure at a nonzero temperature, because of the statistical distribution of velocities, some atoms would pass beyond the inflection point. This approach was verified by molecular dynamics studies. We also studied the effects of missing atoms and 5/7/7/5 defects. The latter have small effects on strength. Missing atoms have significant effects on strength, but the failure strain also decreases quite a bit, so the computed failure strains then underestimate the experimental results.

The application of atomistic models of active bonds such as carbon must generally be treated with caution. However, the great insensitivity of the results to dissociation energies provides some confidence that these models can be used for fracture studies of nanotubes in a variety of situations.

ACKNOWLEDGEMENTS

The support of the Army Research Office, National Science Foundation and the Naval Research Group are gratefully acknowledged by the first two authors. G. Schatz acknowledges the supported of AFOSR MURI grant F49620-01-1-0335. R. Ruoff acknowledges the support of NASA.